\documentclass[superscriptaddress,nobibnotes,prd,nofootinbib]{revtex4}
\usepackage{amsmath}

\usepackage{amsfonts}
\usepackage{amssymb}
\usepackage{graphicx}
\makeatletter
\newcommand*{\rom}[1]{\expandafter\@slowromancap\romannumeral #1@}
\makeatother

\usepackage{dcolumn}
\usepackage{bm}
\usepackage{color}
\usepackage[normalem]{ulem}
\usepackage[percent]{overpic}
\usepackage{natbib}
\usepackage[usenames,dvipsnames]{xcolor}
\usepackage{hyperref}
\hypersetup{colorlinks=true,
citecolor=MidnightBlue, linkcolor=MidnightBlue, urlcolor=MidnightBlue}
\usepackage{tikz}
\definecolor{purple}{rgb}{1,0,1}
\definecolor{lime}{HTML}{A6CE39} 

\usepackage{epsfig}

\newcommand{\orcidicon}{%
	\begin{tikzpicture}
	\draw[lime, fill=lime] (0,0)
		circle [radius=0.16]
		node[white] {{\fontfamily{qag}\selectfont \tiny ID}};
	\draw[white, fill=white] (-0.0625,0.095)
		circle [radius=0.007];
	\end{tikzpicture}	\hspace{-2mm}
}
\newcommand\orcidMarziyeh{{\href{https://orcid.org/0000-0003-2736-9396}{\orcidicon}}}

\begin{document}


\title{Dynamics of kink train solutions in deformed Multiple sine-Gordon models}
\author{Marzieh Peyravi\orcidMarziyeh\!\!}
\email{marziyeh.peyravi@stu-mail.um.ac.ir}
\affiliation{Department of Physics, School of Sciences, Ferdowsi University
of Mashhad, Mashhad 91775-1436, Iran}
\author{Nematollah Riazi}
\email{n\_riazi@sbu.ac.ir}
\affiliation{Physics Department, Shahid Beheshti University, Evin, Tehran 19839, Iran}
\author{Kurosh Javidan}
\email{javidan@um.ac.ir}
\affiliation{Department of Physics, School of Sciences, Ferdowsi University
of Mashhad, Mashhad 91775-1436, Iran}

\begin{abstract}
This paper examines the effects of a thin layer of inhomogeneity on periodic solutions of the Multiple-sine-Gordon (MsG) model. We investigate the dynamics of the perturbed Double-sine-Gordon (DsG) system as a significant and more practical case of such configurations. 
The thin barrier acts as a potential well (potential barrier) and causes critical deformations in kink train solutions and some basic properties of the periodic solutions, such as the type of sub-kinks, their amplitude, energy and wavelength. Stability of the initial kink chain during the interaction with medium defects is analyzed using their phase diagram. Sudden changes in the profile of kink trains due to the disruption of their amplitude and wavelength are considered. The time evolution of moving kink chain solutions while interacting with medium fractures is also studied.

\medskip

{\sc keywords:} Double-sine-Gordon, Multiple-sine-Gordon, Soliton collision, Integrability.

\end{abstract}


\maketitle

\section{Introduction}\label{intro} 

The formation, stabilization, and propagation of localized waves in non-linear media such as solitary waves, solitons, and breathers are fundamental problems in many branches of science. Scattering of solitons from point-like \cite{int1, int2, int3} or extended \cite{int4, int5} defects and impurities as a fundamental problem of soliton theory has been widely investigated. Indeed, such effects play a crucial role in determining the essential physical properties of non-linear systems, and this issue has attracted many theoretical and practical interests \cite{int6, int7, int8}. The processes of soliton trains in quantum optics \cite{soltrain1, soltrain2}, a chain of localized solutions in Josephson-Junction Arrays \cite{soltrain3, fx2}, and the dynamics of soliton ratchets in perturbed background medium \cite{12, 13, 14} are important topics, both in theory and also applications. It is clear from an experimental point of view, we have to consider spatial defects to find a realistic perspective of the system. Inhomogeneity can arise in a medium due to various defects, such as dislocations, impurities, imperfect grain boundaries, etc. Thus, one can see defects as local deviations from an ordered structure in the system \cite{int4, int9, int10}.

The family of Sine-Gordon (SG) models including its modifications, have appeared in different branches of physics and engineering. There are a lot of investigations that focus on multi-sine-Gordon (MsG) models and especially the double-sine-Gordon (DsG) equation. The DsG model has been used to explain the non-linear nature of many physical phenomena like the spin dynamics in the B phase of super-fluid 3He \cite{2,3}, charge density wave condensate models of the organic linear conductors, and dynamics of the liquid crystals \cite{org, lq1, lq2}, quark confinement \cite{10}, self-induced transparency phenomena \cite{11} and many more situations. The non-linear vibrations arising in the engineering applications by considering the noise and uncertain properties \cite{en1}, or propagation of ultra-short optical soliton trains \cite{tr} are modelled by different MsG equations. Family of sG potentials appear in different ways to describe natural non-linear phenomena. The general form of these potentials is in the form $V(\phi)=a+b\cos(n\phi)+c\cos(m\phi)$ , where $a,b,c,n$  and $m$ are system parameters. In practical situations (such as soliton/solitary wave-based telecommunication, applications related to crystal structures and …) one needs soliton train solutions. However non-linear properties of multi-frequency sG potentials have been studied extensively, unfortunately, there exist analytical periodic solutions only for very special cases of such potentials. For this reason, real models are often approximated to such known forms. Due to the non-linear nature of these equations, the applied approximations cannot support all the results of the real phenomenon. In particular, the stability of solutions and their sensitivity respect to variation of model parameters (especially for periodic solutions) in reduced models, have not been investigated. Thus, here we solve the general form of the equations numerically. To ensure the correctness of outcomes, we compare our results (in certain cases) where there are previously published analytical periodic solutions.    
According to our best knowledge, the periodic solution has been found only for the single and double sine-Gordon equations. Unfortunately, such solutions are in series expansion forms. The breather train solution \cite{1j} and periodic solution \cite{2j, 3j} for the standard sG have been presented analytically. The general behaviour of multi sine-Gordon model has been investigated in few papers. One of the best studies is presented in \cite{4j}. The periodic solution for stationary multi sine-Gordon has been investigated numerically in \citep{paper2} by some authors of this manuscript.

In real environments, medium defects in the form of disorders, impurities, or dislocations cause serious changes in the character of propagated waves. The stability of travelling waves, especially in non-linear media is not a simple and straightforward issue. In such problems, we need a phase diagram analysis as well as the frequency–amplitude relationship under the existence of medium inhomogeneities to understand the actual vibration properties of the medium. Investigation of the dynamics of these types of phenomena needs periodic solutions of MsG equations (mostly the DsG model) in the presence of medium defects and impurities. This means that we have to deal with two problems at the same time: finding periodic solutions for MsG equations and also adding medium defects to the model by considering suitable potentials and/or modifications. Motivated by this question, we have studied the interaction of periodic solutions with medium defects.

In this contribution, we consider solutions of the DsG and MsG equation in $1+1$ space-time dimensions. Our goal is to study the effect of some kinds of barriers, somewhere among the chains of the kinks and anti-kinks of the DsG and MsG systems. The amplitude and wavelength of solutions in the presence of different perturbations will be studied. We thus propose to study the modified periodic DsG  and MsG solutions.

The structure of our presentation is as follows:
Section \ref{sec2} belongs to review some general properties of the
DsG and MsG system, including the action, the field equation, and the single kink solution. In Section
\ref{Field} we study the effects of thin barriers on the periodic solutions of the DsG and MsG systems. In
Section \ref{Field1}, we focus on the collision of the DsG with the wall. 
 We close in Section \ref{Field3} by summarizing our
results and pointing out some directions for future work.

\section{Basic Properties of the Double sine-Gordon and the Multiple-sine-Gordon Systems }\label{sec2}
In this section, we briefly review the DsG and MsG formalism by introducing
the related action, the field equation, the energy-momentum tensor, and  the topological current. 

\begin{figure}[ht]
\epsfxsize=9cm\centerline{\hspace{9cm}\epsfbox{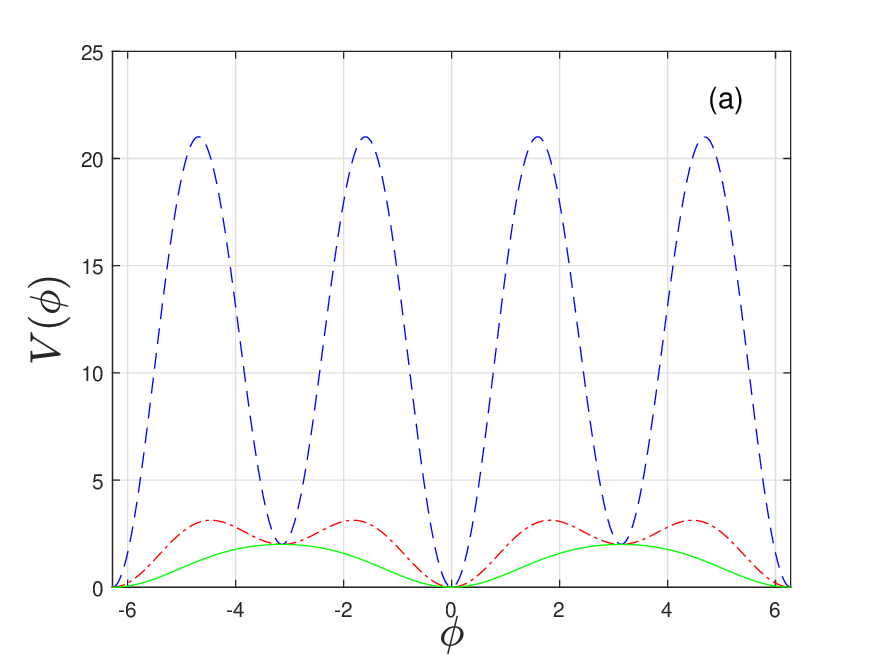}\epsfxsize=9cm\centerline{\hspace{-10cm}\epsfbox{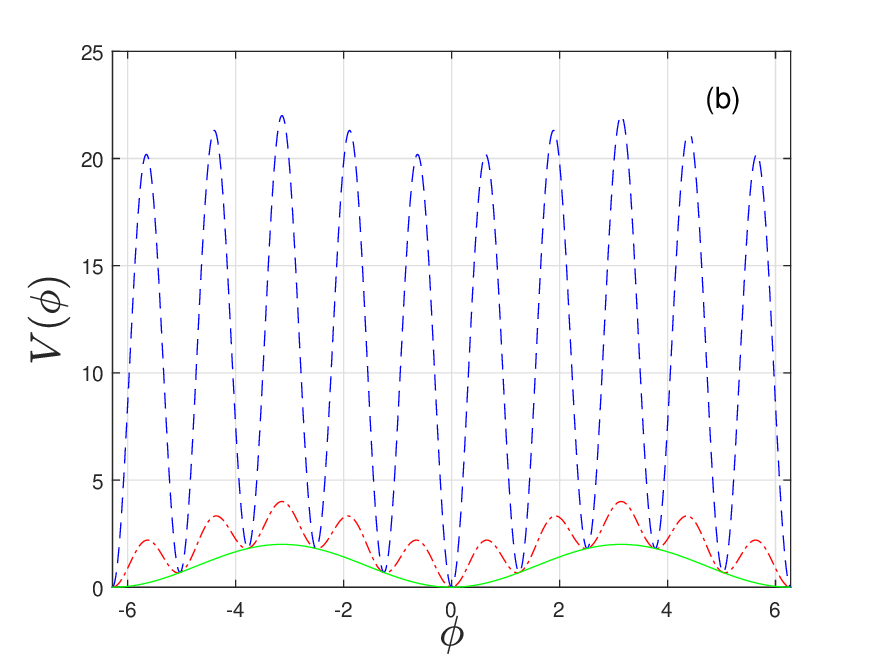}}}
\caption{(a) The DsG Potential. The dashed curve is for $\epsilon=10$, the dash-dotted
curve is for $\epsilon=1$, and the solid curve is for
$\epsilon=0$ (sG). (b) The MsG
Potential for $N=5$. The dashed curve is for $\epsilon=10$, the
dotted curve is for $\epsilon=1$ and the solid curve is for
$\epsilon=0$ .\label{dsgp}}
\end{figure}

Following our previous study on the periodic and step-like solutions of DsG
equation \cite{0} and static properties of the MsG systems \cite{paper2}, we focus on the following potential (see Fig. \ref{dsgp}):
\begin{equation}\label{potten}
V(\phi) = 1 + \epsilon -\cos \phi -\epsilon \cos (N\phi )
\end{equation}
where $\epsilon$ is a non-negative constant and $N$ is an integer with $N=0$, $N=2$, and $N>2$
corresponding to the sG, DsG, and MsG systems, respectively
\cite{0,paper2,ri1}. The harmonic term
in this potential can result from the Fourier expansion of an
arbitrary, periodic potential $\widetilde{V}(\phi) =\widetilde{V}(\phi+2n\pi)$. 
As the Fig. \ref{dsgp} illustrates, the
above potential find its minima at: $\phi_{min}=2n\pi$ with $ n=0,1,2,...^{}$. 
Based on the nature of the potential, more vacua will appear in the profile of the potential, as $N$ increases, and thus, extra sub-kinks (up to $N$) are created in the solution, which leads to a more complicated system \cite{0,paper2}.

While integrable equations are an essential part of the non-linear wave theory,
most non-linear wave equations that emerge in physics and engineering are not integrable \cite{majhool}.
Non-integrable equations involve much richer and more complex dynamical solutions. Non-integrable equations not only lead to unstable solitary waves with complicated and even chaotic interactions via fractal scattering phenomena but also these solitary waves can collapse in more than one dimension \cite{majhool}.

\begin{figure}
\epsfxsize=9cm\centerline{\epsfbox{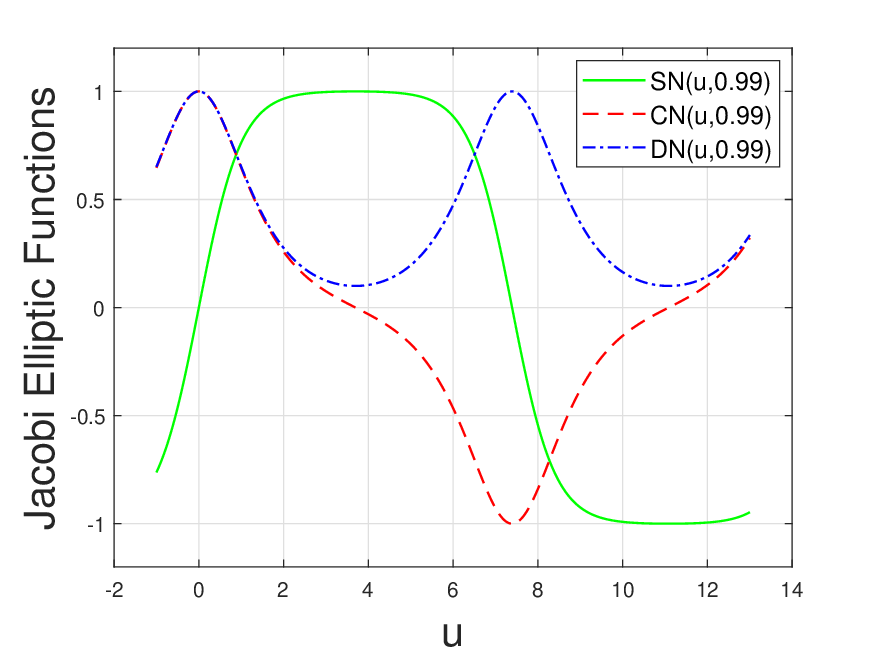}}
\caption{Jacobi Elliptic Functions as a function of u for $k=0.99$.\label{elpff}}
\end{figure}

Several powerful methods have been presented for deriving localized travelling wave solutions for non-linear equations in recent years, for example, the inverse scattering transformation, Hirota bilinear method, Backl\"{u}nd and Darboux transformations, sine-cosine and tanh-function methods, homogeneous balance and Lie group analysis, F-expansion method and so on. Through these methods, solitary, periodic, multiply periodic, quasi- or non-periodic wave solutions as well as kink (anti-kink) trains have been constructed as polynomials in the triangle, hyperbolic, or Jacobian elliptic functions for the DsG model (See Fig. \ref{elpff}) \cite{j.amc, DsGJaccobi, DsG periodic}. Several kink train solutions for the DsG model have been proposed using Jacobi's elliptic functions, such as:
\begin{equation}\label{Sc}
\phi=\pi + 2 \arctan \left[ (k^{\prime}_2) ^{-1}  JN \bigg( \frac{2k^{\prime}_2 \sqrt{\epsilon}}{k_1\sqrt{-k^2_2} } (x-x_0-vt),k   \bigg) \right].
\end{equation}
where, $JN(u,k)$ is a suitable generalized Jacobi elliptic function, related to the potential parameters, $k^2_{1,2}=\frac{1}{3}\left[ (1-4\epsilon)\pm \sqrt{(1+\epsilon)^2+8\epsilon}\right]$, $k^2=\frac{k_1^2-k_2^2}{1-k_2^2}$, and $k^{\prime}_2=\sqrt{1-k_2^2}$ \cite{DsGJa}.
The Jacobi elliptic functions take the parameters $u$ and $k$ as inputs. In the following integration: 
\begin{equation}\label{Jacobi}
u=\int^{\phi}_{0}\frac{d\theta}{\sqrt{1-k sin^2 \theta}}
\end{equation}
the parameter $\phi$ is called the Jacobi amplitude $A(u,k)=\phi$. In this framework, the elliptic sine is introduced as $SN(u,k)=\sin(A(u,k))$ and the elliptic cosine is given by $CN(u,k)=\cos(A(u,k))$. In similar definition we can define the elliptic delta amplitude as $DN(u,k)=\frac{d}{du}\left( A(u,k)\right)$. Figure \ref{elpff} demonstrate These three Jacobi elliptic functions respect to the parameter $u$ with $k=0.99$.  

 Explicit solutions of non-linear equations are very helpful to analyse all the dynamical features of the system. Unfortunately, only periodic solutions of sG and DsG potentials have been obtained using different approaches, which are often not explicit \cite{1jj,2jj}. The periodic solutions of the DsG potentials are presented as expansions of Jacobi functions, which can be converted into closed form only in certain values of model parameters, which are often not physical \cite{3jj}. Explicit solutions for the MsG equation have not been found yet. Thus, we can only rely on numerical calculations. To control the accuracy, the results of numerical calculations on the MsG potential is reduced to the DsG model and then compare them with analytical solutions of the DsG equation. 

According to our best knowledge, there are no such solutions for the general form of the MsG system. Therefore we calculated periodic solutions for the MsG potential numerically. Indeed, most of the presented solutions for more complicated potentials, have been given in infinite series expansions. In such situations, we have to use numerical calculations/simulations to study the behaviour of the solutions. Nowadays, numerical computations play an important and inevitable role in non-linear science and soliton theory \cite{majhool}. Since non-integrable systems are not solvable analytically, numerical methods provide powerful tools for non-linear wave studies.

The Lagrangian density of the system consisting of the MsG system for real scalar field  $\phi(x,t)$ in $(1+1)$ dimensions is in the following form \cite{0,paper2}:
\begin{equation}\label{msg}
{\cal L}=\frac{1}{2} \partial^\mu \phi \partial_\mu \phi
-\left[ 1+\epsilon -\cos\phi -\epsilon \cos (N\phi
)\right].
\end{equation}
From this Lagrangian density, the field equation of motion is as follows \cite{0,paper2}:
\begin{equation}\label{shok}
\Box \phi =-\sin \phi -N\epsilon \sin (N\phi ).
\end{equation}

The energy-momentum tensor of the MsG equation can be obtained by
using Noether's theorem and the invariance of the action under the space-time translation
$x^{\mu}\rightarrow^{\mu}+a^{\mu}$ \cite{0,paper2,22,23}:
\begin{equation}
T^{\mu\nu}=\partial^\mu \phi \partial^\nu \phi -g^{\mu\nu}{\cal L},
\end{equation}
in which $g^{\mu\nu}=diag(1,-1)$ is the metric of the Minkowski
space-time. Besides, a topological current can also be
defined for a generalized form of sG through the following current \cite{24}:
\begin{equation}
J^\mu =\frac{1}{2\pi}\epsilon^{\mu\nu}\partial_\nu \phi,
\end{equation}
where $\epsilon^{\mu\nu}$ is the antisymmetric tensor with
$\epsilon^{01}=1$.  This current is identically conserved
($\partial_{\mu}J^{\mu}=0$) and the total topological charge of any localized,
finite-energy solution is conserved and also quantized. We have used the topological charge and energy of the system for controlling the validity of our numerical calculations in every selected step.

\begin{figure}
\epsfxsize=9cm\centerline{\hspace{9cm}\epsfbox{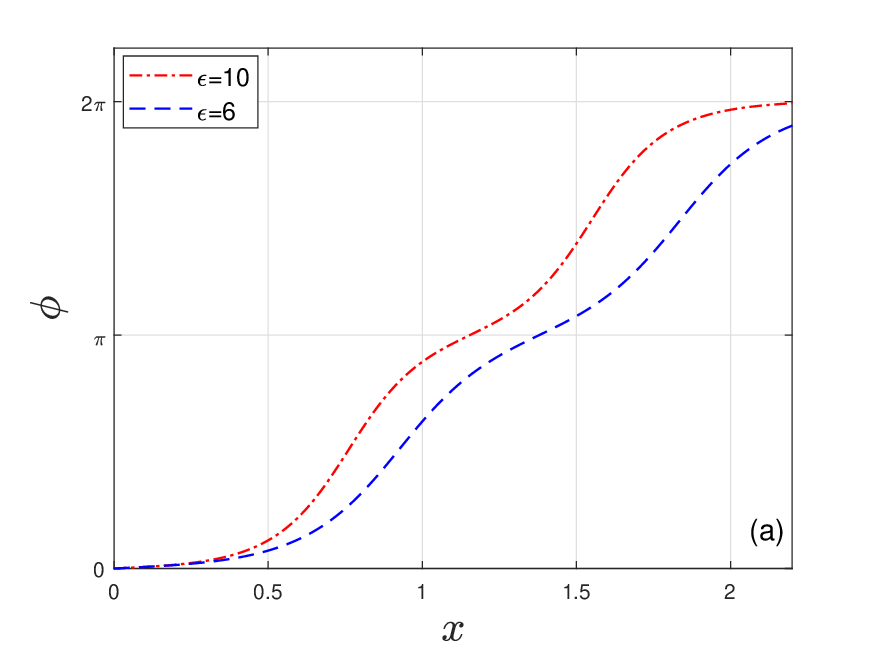}\epsfxsize=9cm\centerline{\hspace{-10cm}\epsfbox{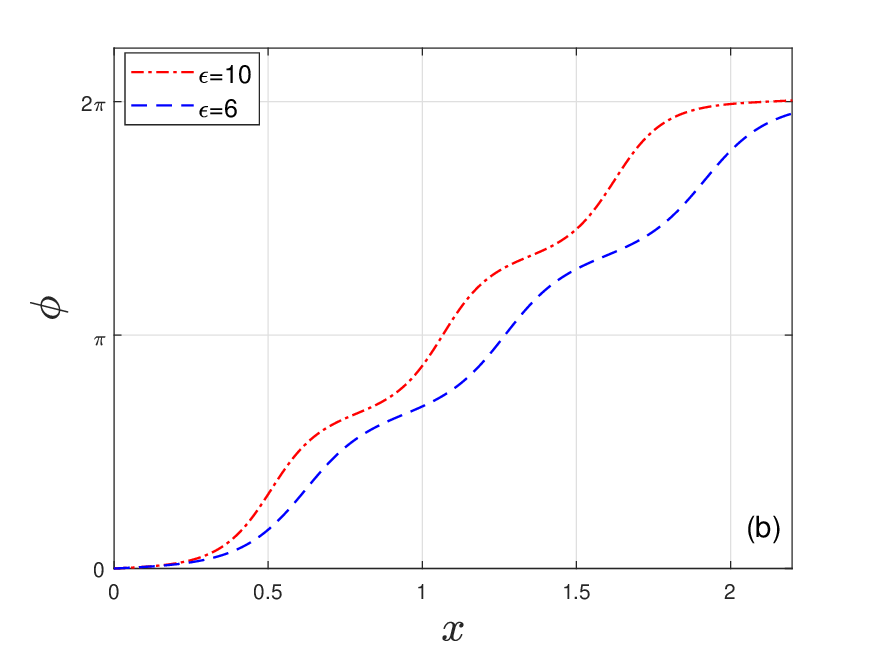}}}
\caption{ Single soliton solutions (a) For DsG with
$\epsilon=10$ (dash-dotted curve) and $\epsilon=6$ (dashed curve) 
(b) The MsG system ($N=3$) with $\epsilon=10$ (dash-dotted curve) and $\epsilon=6$ (dashed curve) . \label{h}}
\end{figure}

As one can check, the MsG equation possesses kink and anti-kink solutions which
correspond to transitions between the spatial boundary conditions
$\phi (\pm \infty )=2n\pi$ \cite{0,paper2}. The first integral of the equation
(\ref{shok}) for static, localized solutions reads:
\begin{equation}\label{kinkp}
\frac{1}{2}(\frac{d\phi}{dx})^2 =V(\phi),
\end{equation}
in which we have used the boundary conditions $\phi (\pm \infty
)=2n\pi$. We therefore have:
\begin{equation}
x-x_0=\int \frac{d\phi}{\sqrt{2V(\phi )}}.
\end{equation}
Multiple sine-Gordon systems admit soliton-like kink solutions
with interesting properties, although this integral cannot be carried out analytically for general $N$
and $\epsilon$ specially for $N>2$. However for $N=2$ one can show that the following exact, static,
single kink (anti-kink) solution exists \cite{0,ri1,paper2}:
\begin{eqnarray}\label{b}
\phi(x) = 2\arccos\left[\pm
\frac{\sinh\sqrt{4\epsilon+1}x}{\sqrt{4\epsilon+\cosh^{2}\sqrt{4\epsilon+1}x}}\right].
\end{eqnarray}
The $(+)$ and $(-)$ sign are correspond to the anti-kink ($\tilde{k}$) and the kink ($k$) solutions respectively \cite{0,ri1}. 
Figs. \ref{h} illustrate DsG kink/anti-kink Eq. (\ref{b}) for two values of the parameter $\epsilon(=6, 10)$ as well as solutions of $N=3$  which is obtained numerically. We can interpret and adjust our numerical solution for the kink train in DsG by comparing it with the exact solution (\ref{Sc}).

It is easy to show that solutions of the MsG equation are classified and characterized by  (x-independent $\frac{dP}{dx}=0$) constant of integration $P$:
\begin{equation}\label{222}
P=\frac{1}{2}\left(\frac{d\phi}{dx}\right)^2-V(\phi).
\end{equation}
in two types, namely: Step-like and periodic solutions (See Fig. \ref{AB}).
Note that this
classification is independent of $\epsilon$ and therefore holds true also for sG equation. The step-like solutions are a sequence of kinks ($kk$) or anti-kinks ($\tilde{k}\tilde{k}$) which are characterized by $P>0$, and the periodic solutions are a sequence of kink, anti-kink
($k\tilde{k}k\tilde{k}$) solutions that are characterized for $-2<P<0$ \cite{0,paper2}. The general behaviour of our numerical solutions is in agreement with reported solutions in Ref. \cite{baz}. Recently, a set of new kink solutions for special types of MsG models have been presented, using the mentioned Jacobi elliptic functions \cite{joseph}. In such cases, our static solutions are successfully reduced to kink solutions in the Ref. \cite{joseph}. Also, there are localized solutions for the triple sine-Gordon equations ($N=3$) for specific values of the model parameters \cite{TsG}.

In fact, one can interpret $P$ as the ``pressure'' (or tension in this 1d case) when we consider such systems as a many-body interacting system \cite{Moradi.EPJB.2018,Gani.EPJC.2018,Gani.EPJC.2019}.
Note that $P$ is different from energy density
\begin{equation}\label{8}
\rho(x)=\frac{1}{2}\left(\frac{d\phi}{dx}\right)^2+V(\phi);
\end{equation}
which changes with position $x$ \cite{0,paper2, Moradi.EPJB.2018}. However MsG potential contains infinite topological sectors, but the multiplicity of solutions is equal to $N$ which are different in terms of the number of their sub-kinks. \cite{baz}. In addition to soliton train solutions, the model contains breather solutions for $N^2 < 4  \pi$ \cite {bth}.

\section{The effect of thin barriers }\label{Field}

In this section, we report the behaviour of kink train solutions of the DsG equation in the
presence of a thin interval of inhomogeneity. For this
purpose $\epsilon$ is considered to be different within a finite
range $x_{1}<x<x_{2}$ in which $x_1$ and $x_2$ are known locations while $x_2 > x_1$. 

As stated before, the behaviour of the solution depends on the value of parameter $P$, which is a function of field tension (stress) $\frac{d\phi}{dx}$ and also the potential ($U(\phi)$). This means that we can control the field solution by correctly choosing the value of the field ($\phi(x)$) and its spatial derivative ($\frac{d\phi}{dx}$) as initial conditions. It is clear that medium dislocations and impurities change these conditions, and therefore such disorders critically affect the field solution in the medium.\\
The area of perturbation, where the barrier-field interaction starts ($x_1$), and terminates ($x_2$), barrier width ($x_2-x_1$), and its amplitude ($\epsilon$) are important parameters. These parameters determine the general characteristics of the solution in the perturbed area as well as the kink train configuration after passing through the barrier.

\begin{figure}[ht]
\epsfxsize=9cm\centerline{\hspace{9cm}\epsfbox{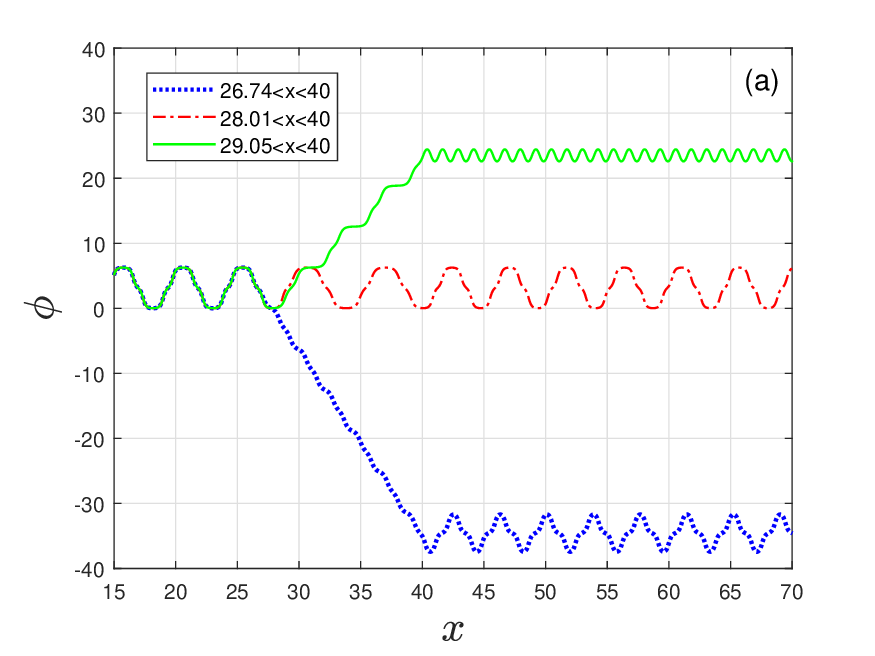}\epsfxsize=9cm\centerline{\hspace{-10cm}\epsfbox{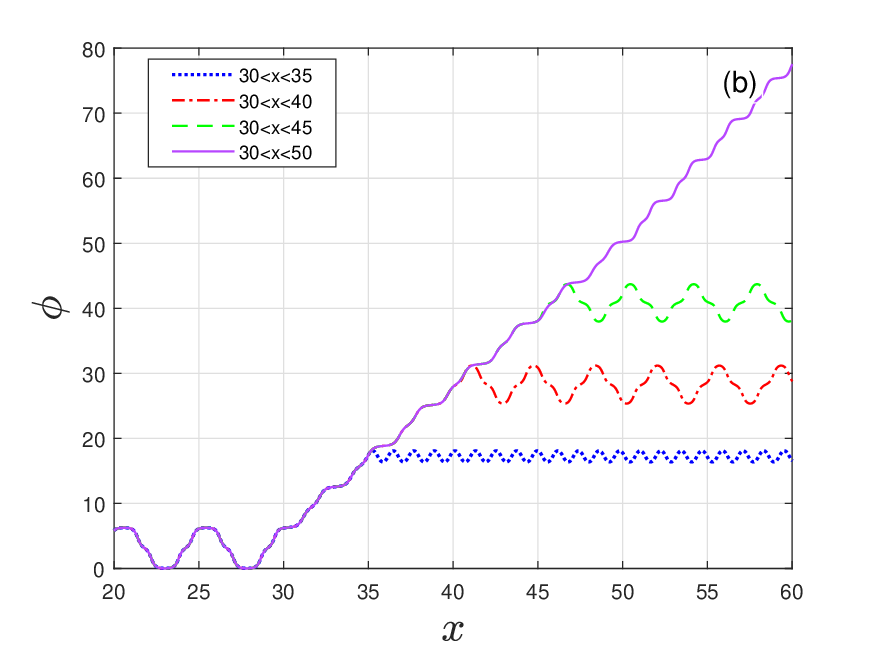}}}
\caption{(a) $\phi$  diagram as a function of $x$ for $N=2$ and $P=  -0.009987500000000$; $\epsilon=10$ for $0 \leq x \leq x_{1}$, $x_{2} \leq x \leq 100$ and $\epsilon=5$ for $x_{1}<x<x_{2}$. For dotted, dash-dotted and solid curves $x_{1}=26.74, 28.01, 29.05$  and $x_{2}=40$ respectively. (b) $\phi$ diagram as a function of $x$ for $N=2$ and $P= -0.009987500000000$; $\epsilon=10$ for $0 \leq x \leq x_{1}$, $x_{2} \leq x \leq 100$ and $\epsilon=5$ for $x_{1}<x<x_{2}$. For dotted, dash-dotted, dashed and solid curves  $x_{1}=30$ and $x_{2}$  are corresponded to $35, 40, 45$ and $50$ respectively. \label{AB}}
\end{figure}

\begin{figure}[ht]
\epsfxsize=9cm\centerline{\epsfbox{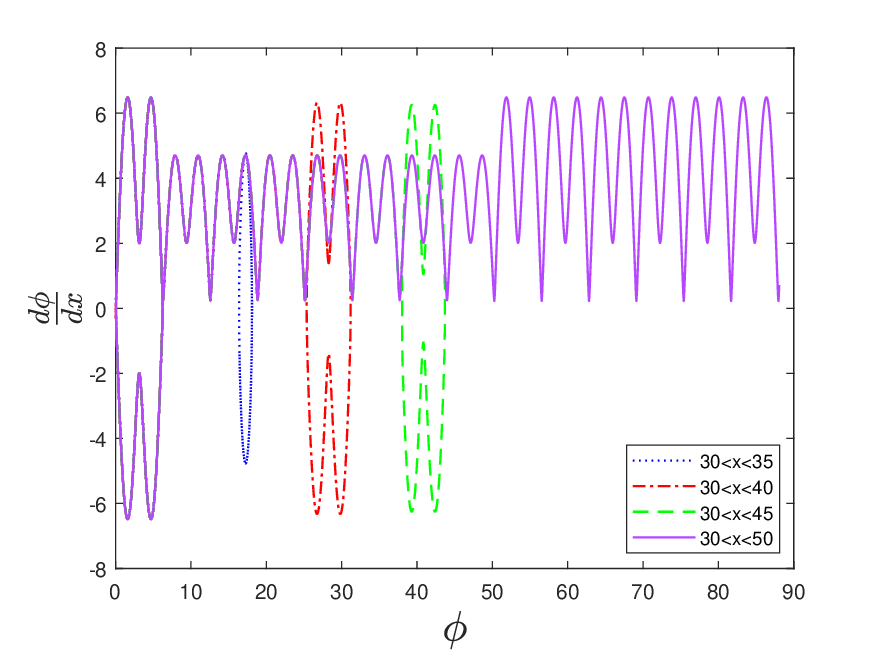}}
\caption{ The phase diagram for $N=2$ and $P= -0.009987500000000$; $\epsilon=10$ for $0 \leq x \leq 30$, $x_{2} \leq x \leq 100$ and $\epsilon=5$ for $x_{1}<x<x_{2}$. Dotted, dash-dotted, dashed and solid curves $x_{2}$  are corresponded to $35, 40, 45$ and $50$ respectively. \label{ABb}}
\end{figure}

The minimum and maximum value of parameters  depend on the possible values of $\frac{d\phi}{dx}$  and 
$V(\phi)$ as one can find from Eq. (\ref{222}). These values generally depend on the condition of the field at the starting point of the perturbation area and the characteristics of the barrier. The constraints imposed at the border points of the barrier can limit the possible values after the kink train leaves the perturbation region. The set of initial conditions for Fig. \ref{AB}(b) causes the response to appear as a kink train in the perturbation region. The possible values for the $\frac{d\phi}{dx}$ are always greater than the value of the potential at the barrier $V(\phi(x))$  (see Fig. \ref{ABb}). Indeed, by changing the length of the perturbation region, it is not possible to decrease the value of P to find anti-kink train solution after the interaction. It is clear that by changing the characteristics of the barrier as well as initial kink train solution, other different train solutions may occur.

 Kink and anti-kink solutions Eq. (\ref{b}) are opposite to each other in terms of spatial expansion. Thus, the field of a kink train ($kk$) solution is an increasing function with respect to the space. The field solution of anti-kink train ($\tilde{k}\tilde{k}$) is a spatial decreasing function, but the solutions in the form of  kink anti-kink trains ($k\tilde{k} $) are periodic functions of the space coordinate. One can find all these configurations in Fig. \ref{AB}. The initial configuration at $x<x_1$ of the  Fig. \ref{AB}(a) is a  $k\tilde{k}$ solution with a periodic variation. The dotted line curve in the Fig. \ref{AB}(a)  demonstrates a ($\tilde{k}\tilde{k}$) configuration, while the solid line shows a ($kk$) kink train solution. 

Figure \ref{AB}(a) shows critical changes related to the starting point of perturbation ($x_1$). By a slightly changes 
of $x_{1}$, we can provide step-like kink ($kk$) (anti-kink $\tilde{k}\tilde{k}$) trains or periodic $k\tilde{k}$
configurations, in the perturbation region. It is due to changing the initial condition for creating the kink solution in the perturbed region.

The most important point is related to the slope of arriving kink (anti-kink) as interacted with the perturbation. Fig. \ref{AB}(a), clearly shows that the kink (anti-kink) train is created in the perturbation region if the initial kink arrives with a positive (negative) slope. If the initial kink arrives with a very small slope, we will find $k\tilde{k}$ configuration in the perturbed region. 

Figure \ref{AB}(b) demonstrates the interaction of kink solution with perturbations starting from the same initial position (here $x_{1}=30$ with $\epsilon=5$) but with different widths. Emerging configuration from the perturbed region depends on the characters of the solution at the end point of the perturbation region ($x_{2}$). In general, the final state of the solution after interacting with the barrier can be any of the possible solutions. Some kink configurations may not occur depending on the possible values for the slope of the field in the barrier. In Figure \ref{AB}(b), we do not expect the formation of the state ($\tilde{k}\tilde{k}$), because the $kk$ configuration which has been established in the perturbation region, does not have a negative slope. Thus, the final state of the chain configuration after passing the barrier can be periodic ($k\tilde{k}$ ) or step-like ($kk$).  We may observe the emergence of $k\tilde{k}$  configuration if the kink reaches the end point of the perturbation area with its flat tail (Please see final $k\tilde{k}$ configurations in the Fig.  \ref{AB}(b) for $x_{2}=35, 40$ and $45$). On the other hand, a kink train configuration may occur after the interaction, if the field solution reaches the barrier endpoint with its positive slope parts. Similarly, if the established configuration in the perturbed area does not have a positive slope, the $kk$ train will not occur after the field-impurity interaction.

Our other simulations are in agreement with this procedure. If the anti-kink configuration is created within the barrier region, we can reach $k\tilde{k}$ or $\tilde{k}\tilde{k}$ trains. For a $k\tilde{k}$ configuration in the barrier, we expect  to find $k\tilde{k}$, $kk$ or $\tilde{k}\tilde{k}$ trains according to the final state of solution where the soliton-barrier interaction is completed.
The potential barrier may be an impurity/spatial dislocation or a defect in the crystal lattice, where the kink train propagates in real physical situation. According to the Figure \ref{AB}, due to a sudden change in the texture of the background space, the characteristic of the kink train experiences critical changes. As Figure \ref{AB} shows, the interaction of the kink train with a structural fault causes critical changes in the topological characteristics of the solution. Of course, we have to pay attention on the dynamical behaviour of the system, such as stability of solution, kink train shape, its velocity after collision. To understand such these characters, we study the description of the barrier-field interaction in the phase space as shown in the Fig. \ref{ABb}.  The quantity $\frac{d\phi}{dx}$ actually refers to the field stress through the term $\frac{1}{2}(\frac{d\phi}{dx})^2$ which expresses the amount of exerted force by the field (per unit length, in $1+1$ dimensions). 

 Figure \ref{ABb} demonstrates the phase space diagram, for configurations of Fig. \ref{AB}(b).  As one can see, for the shortest barrier ($30<x<35$, the dotted curve) phase diagram has only one ring due to the absence of sub-kinks. For different impurity widths ($30<x<40$ and $30<x<45$, dash-dotted, dashed curves) we find other diagrams containing twofold closed loops due to the presence of sub-kinks. It should be noted that these diagrams relate to the same type of solution but in different sectors. The noticeable point in the Fig. \ref{ABb}, relates to the solid open curve which corresponds to the $kk$ train with sub-kinks.
We learn from Fig. \ref{ABb} that, the footprint of each kink (anti-kink) is a single or double concave (convex)-like curve in the phase space diagram. $k\tilde{k}$ configurations create closed loops in a limited area of the phase space plane if both kink and anti-kink are located in the same sector.
But $kk$ and $\tilde{k}\tilde{k}$ configurations are not closed curves. This means that, each kink (anti-kink) in created $kk$ or $\tilde{k}\tilde{k}$ trains are not located in the same sector. However the multiplicity of solutions in this model is $N=2$, but topological charge of all kinks (anti-kinks) are the same. Thus, all created objects in a train solution, show similar behaviour \cite{baz}.

\begin{figure}[ht]
\epsfxsize=9cm\centerline{\hspace{9cm}\epsfbox{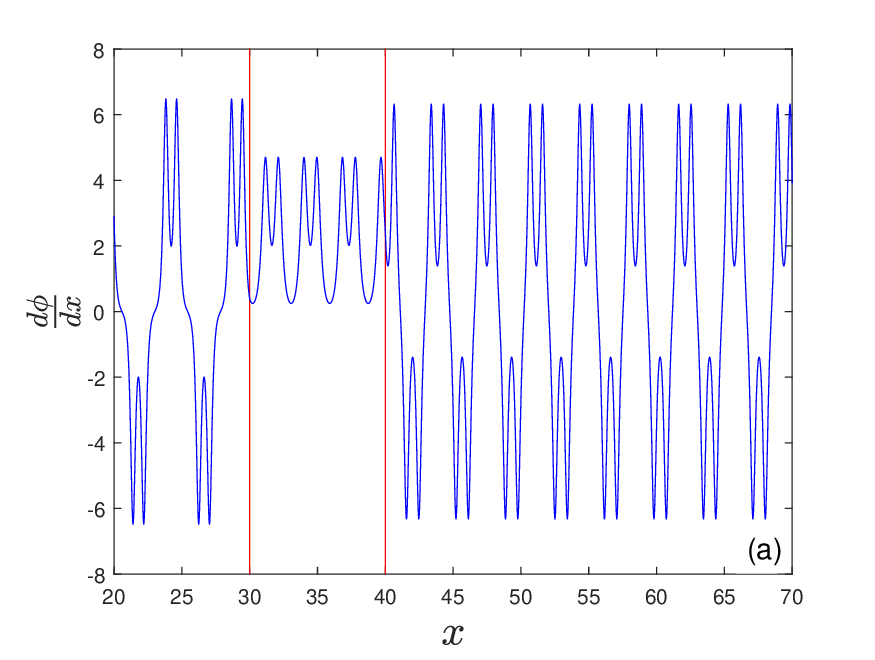}\epsfxsize=9cm\centerline{\hspace{-10cm}\epsfbox{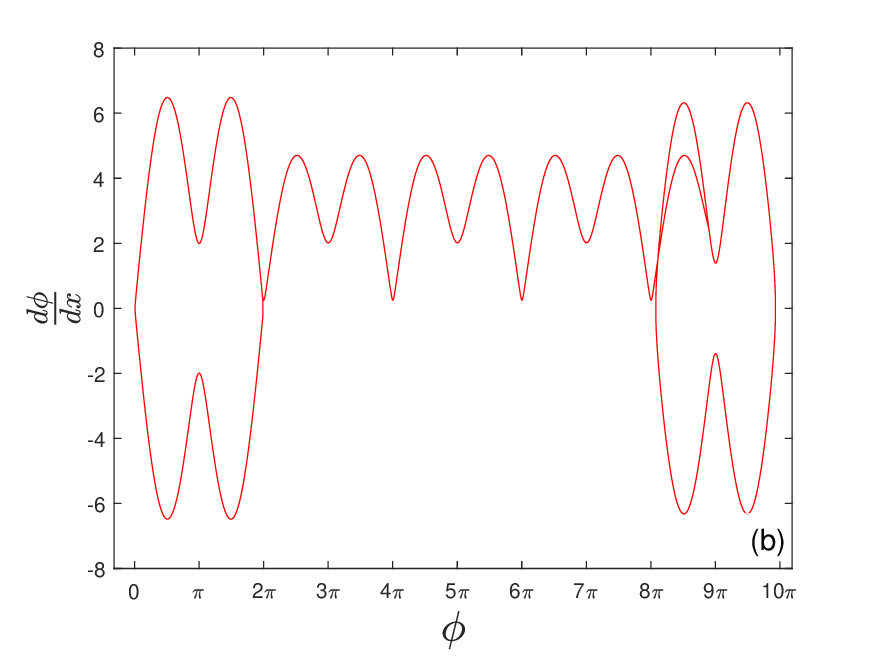}}}
\caption{(a) The $\frac{d\phi}{dx}$  diagrams as a function of $x$ and (b) The phase diagram $N=2$ and $P=  -0.009987500000000$; $\epsilon=10$ for $0 \leq x \leq30$ and
$40 \leq x \leq100$ and $\epsilon=5$ for $30<x<40$. \label{1}} 
\end{figure}

\begin{figure}[ht]
\epsfxsize=9.1cm\centerline{\hspace{9cm}\epsfbox{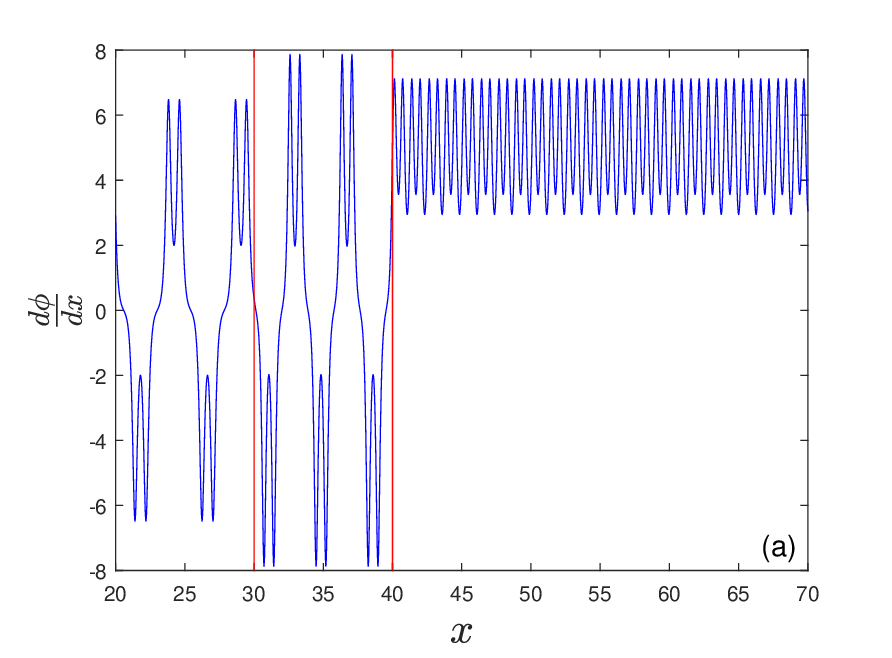}\epsfxsize=9.2cm\centerline{\hspace{-10cm}\epsfbox{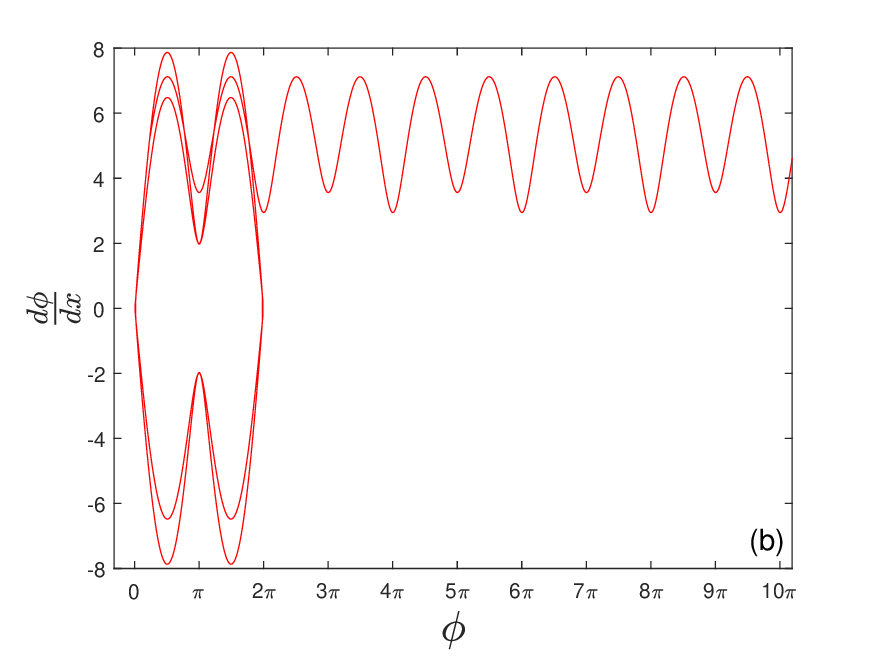}}}
\caption{(a) The $\frac{d\phi}{dx}$ diagram as a function of $x$ and (b) The phase
diagram $N=2$ and $P=-0.009987500000000$; $\epsilon=10$ for $0 \leq x \leq30$ and
$40 \leq x \leq100$ and $\epsilon=15$ for $30<x<40$. \label{2}}
\end{figure}

\begin{figure}[ht]
\epsfxsize=9cm\centerline{\hspace{9cm}\epsfbox{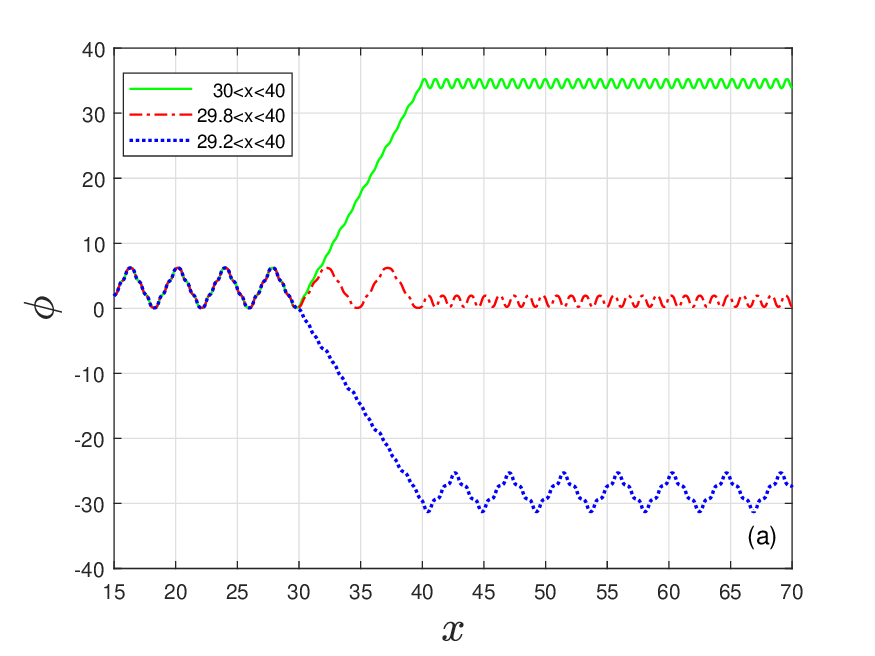}\epsfxsize=9cm\centerline{\hspace{-10cm}\epsfbox{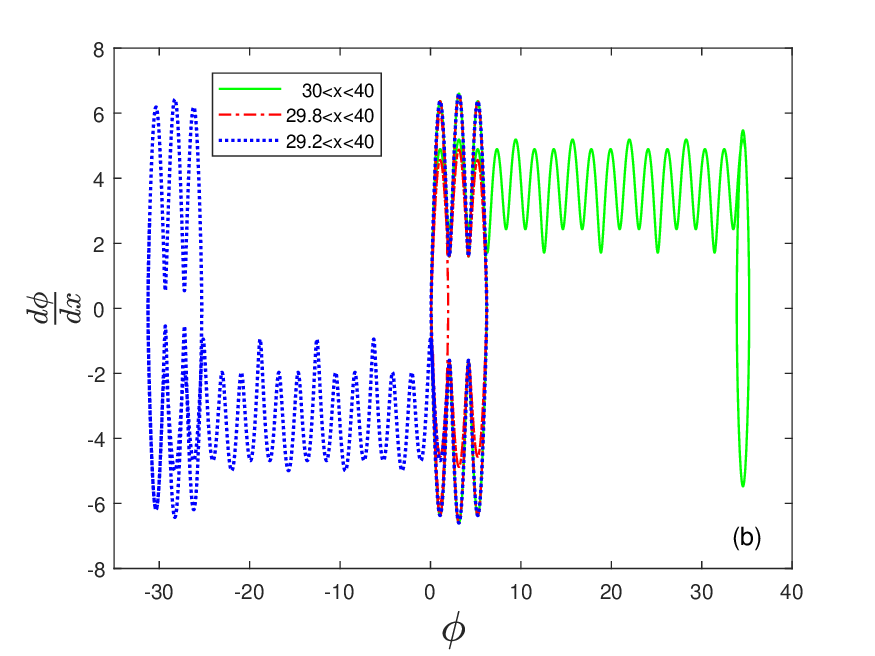}}}
\caption{(a) $\phi$  diagram as a function of $x$ and (b) The phase diagram for $N=3$ and $P= -0.220000000000000$; $\epsilon=10$ for $0 \leq x \leq x_{1}$, $x_{2} \leq x \leq 100$ and $\epsilon=5$ for $x_{1}<x<x_{2}$. For dotted,  dash-dotted and solid curves $x_{1}=29.2, 29.8, 30$  and $x_{2}=40$ respectively. \label{AB2}}
\end{figure}

Dynamic behaviours depending on open and closed paths in the phase space can be explained based on equation (\ref{8}). The energy of a kink chain configuration can be obtained by calculating potential energy $V(\phi)$ and stress $\frac{1}{2}(\frac{d\phi}{dx})^2$ . Figure \ref{dsgp} shows that function $V(\phi)$ is a periodic function, and always changes in a limited range. In other words, this quantity never tends to infinity. But the $\frac{d\phi}{dx}$  (at least theoretically) has no limit. Phase space ($\phi-\frac{d\phi}{dx}$) clearly demonstrates the variation of $\frac{d\phi}{dx}$ in terms of the field solution. Phase space paths in figures \ref{ABb}-\ref{AB2} demonstrate the changing range of $\frac{d\phi}{dx}$ in terms of field value $(\phi)$. The phase space curves represent the exchange of energy between its two different forms: potential energy $V(\phi)$  and stress $\frac{1}{2}(\frac{d\phi}{dx})^2$. The closed paths in the phase space correspond to the ($k\tilde{k}$) solutions, while the open paths regard to ($kk$) or ($\tilde{k}\tilde{k}$)  train solutions. The phase space diagram of Fig. \ref{1} clearly shows that the amplitude of $\frac{d\phi}{dx}$ changes during the interaction with barrier. As a result, the energy of the incoming solution changes due to collision with the perturbation. But the energy of the kink train returns into its initial value after interaction. This scenario does not occur in all interactions. For example, Fig. \ref{2} shows that the energy of the related kink train changes during the interaction with the perturbation, but it is not restored to its initial value after leaving the barrier.

For finding better knowledge about the similarity and differences between created kinks (anti-kinks) during the field-impurity interaction, we have plotted Figs. \ref{1} and \ref{2} with more details. In these figures, similar $k\tilde{k}$ trains hit  the barriers, which are located in the same spatial position, but with different values of $\epsilon=5$ (Fig. \ref{1}) and $\epsilon=15$ (Fig. \ref{2}). Initial conditions for both situations are the same, but the related barrier potential of defined situations is different. For barrier with $\epsilon=5$ a $kk$ chain is created in the perturbed area. The solution for barrier with $\epsilon=15$ is a $k\tilde{k}$ train with different energy but in the same sector ($0\le \phi \le 2\pi$) as compared with initial $k\tilde{k}$ (see Fig. \ref{1}). One can find from the phase space diagrams in Figs. \ref{1} and \ref{2} that, both solutions in the perturbation region have positive, zero, and negative slope parts. Thus, all types of kink chain solutions can be created after the interaction with defect ($x\ge x_2$), according to the $\frac{d\phi}{dx}$ at the endpoint of perturbation ($x_2$).

These figures show that the established solutions for these two scenarios are different. Indeed $kk$ train ($k\tilde{k}$ solution) is created in the perturbation region for $\epsilon=5$ ($\epsilon=15$). According to the phase space diagram of the Fig. \ref{1} (Fig. \ref{2}), created $k\tilde{k}$ solution after (during) the field-impurity interaction is located in the different (similar) sector as compared with initial $k\tilde{k}$ train. 
An interesting issue is the creation of kinks that are located in different spatial solutions but in similar topological sectors. According to the Fig. \ref{1} (Fig. \ref{2}), kinks which are located in the regions $x_1 \le x \le x_2$ and $x \ge x_2$ ($x \le x_1$) are related to the same sector $8 \pi \le \phi \le 10\pi$ ($0 \le \phi \le 2\pi$).  Figure \ref{2} demonstrates the collision of the kink train with a spatial defect (potential well) with lower density (smaller $\epsilon$). It can be clearly found that the stress decreases in the thinner medium. A very interesting point is that the amount of stress density returns to its initial situation (with a good approximation) after the termination of the field-impurity interaction as one can find from Fig. \ref{1}(b).

 Moreover, it is vital to emphasize the role of $\epsilon$ value via the appearance of potential well/barrier in the medium, in addition to the initial position of perturbation. As the Figs. \ref{AB}-\ref{2} demonstrate, the $\epsilon$ value in the perturbed area has a decisive effect on the slope of  $kk$ or $\tilde{k}\tilde{k}$ train in the barrier region, which leads to the emergence of different $k\tilde{k}$ chains after the interaction with the barrier. In other words, the amplitude, and wavelength of the $k\tilde{k}$ chain will change as a consequence of sub-kinks.


In our simulations, $P$ is calculated numerically by solving
the relevant differential equation. Indeed, we need to know how the $P$ value changes abruptly across the barrier. Here, we try to obtain an analytical expression for the change in $P$ across the
potential ($\bigtriangleup P$). We obtain our results for
time-independent static solutions, however, they can be easily
generalized to time-dependent solutions.

 Consider two regions of different $\epsilon$ joining each other
 at $x=0$. Integrating the static equation from $x=-\delta$ to $x=+\delta$

\begin{equation}
\int^{+\delta}_{-\delta}\frac{d^2\phi}{dx^2}dx=\int^{+\delta}_{-\delta}\frac{\partial
V(\phi)}{\partial \phi}dx,
\end{equation}
and letting $\delta\longrightarrow0$, we obtain

\begin{equation}
\frac{d\phi}{dx}|_{0^{-}}=\frac{d\phi}{dx}|_{0^{+}}
\end{equation}
since there is no singularity in the potential around $x=0$.
Together with the continuity of $\phi$, we are now able to obtain
the relation between the $P$-values on either side of the
junction:

\begin{equation}
P_{1}=\frac{1}{2}(\frac{d\phi}{dx})^2-V_{1};  \qquad
\text{for $x<0$,}
\end{equation}
and
\begin{equation}
P_{2}=\frac{1}{2}(\frac{d\phi}{dx})^2-V_{2};  \qquad
\text{for $x>0$,}
\end{equation}
 using the continuity of $\phi$ and $\frac{d\phi}{dx}$
 across the junction, we obtain
\begin{equation}\label{pp}
P_{2}-P_{1}=V_{1}-V_{2}=\left(\epsilon_{1}-\epsilon_{2}\right)\left(1-\cos(2\phi(0))\right);
\end{equation}
for the DsG equation, when $\phi(0)$ is the value of $\phi$ at
$x=0$(junction point). One observes that even for
$\epsilon_{2}\neq\epsilon_{1}$, we can still have
$P_{2}=P_{1}$, i. e., if $\phi(x=0)=n\pi$. Otherwise,
the value of $P$ will change abruptly across the junction.
The minimum and maximum change in the value of $P$ is 0 and
$2|\epsilon_{1}-\epsilon_{2}|$, respectively. Similar relations can be presented for the MsG models too.

\begin{figure}
\epsfxsize=9cm\centerline{\epsfbox{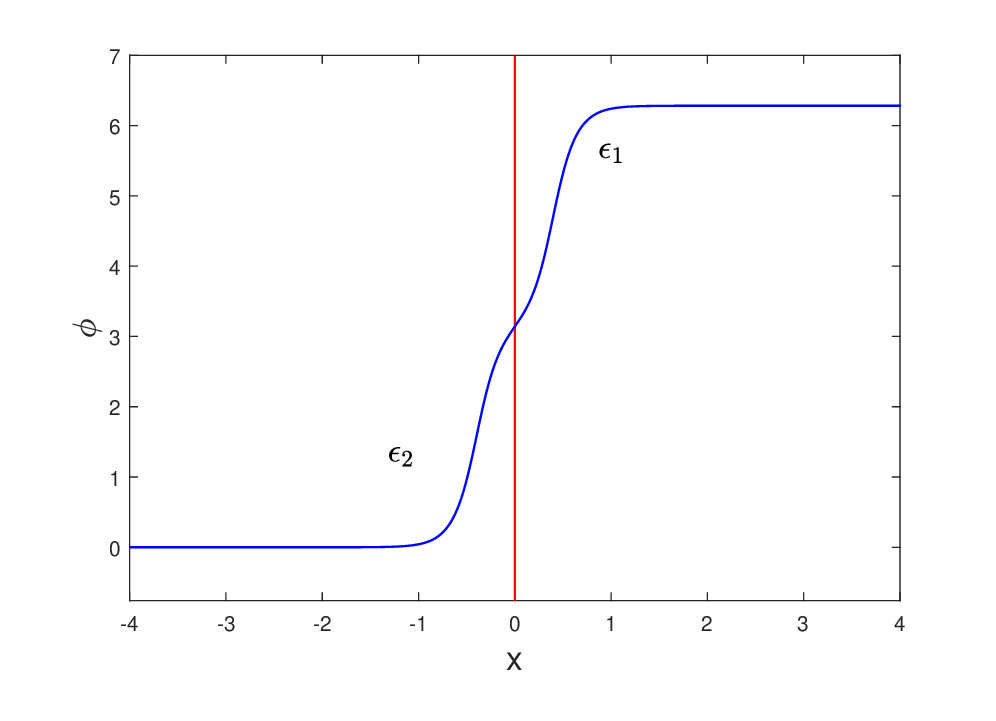}}\caption{A single soliton across a boundary at $x=0$.\label{e1e22}}
\end{figure}

For a single soliton, we have $P=0$, if we consider a media with two different $\epsilon$ on both sides of the boundary (on the left $\epsilon_{2}$ and the right $\epsilon_{1}$) \cite{paper2}. Suppose that there is a single soliton on the left side of the boundary ($P_{2}=0$), so the Eq.(\ref{pp}) reduces to:
\begin{equation}\label{pp1}
P_{1}=\left(\epsilon_{2}-\epsilon_{1}\right)\left(1-\cos(2\phi(0))\right);
\end{equation}
In this case, it is obvious that the vital condition for having the single soliton on the right side of the boundary is $\epsilon_{2}=\epsilon_{1}$ or $\phi_{0}=n\pi$ (See Fig. \ref{e1e22}).

According to Eq. (\ref{pp}), the difference value of the parameter $P$ while entering the perturbation area ($x=x_{1}^{+}$) becomes $P_1=P_2-(\epsilon_1-\epsilon_2)V(\phi(x_1))$ , while $\epsilon_1$ ( $\epsilon_2$ ) relates to the $x>x_1$  ($x<x_1$). We have chosen $\epsilon_2=10$ ($x<x_1$) and $\epsilon_1=5$ ( $x>x_1$ ) in the Fig. \ref{AB}. Therefore, $\Delta P_{x_1}=+5V(\phi(x_1))$ (at starting point) and $\Delta P_{x_2}=-5V(\phi(x_2))$ (while kink train leaves the barrier). Since $V(\phi(x))$ is always positive, the value of parameter $P$ increases at the starting point of the barrier, while it will decrease at the end of perturbed region. To show the effect of $P$ in the configuration of kink train while interacting with the barrier, we have chosen the starting point of the boundary ($x_1$) to be close to the minimum value of $V(\phi(x))$ for plotting the Fig. \ref{AB}(a). In this case, for an initial value (in the Fig. \ref{AB}(a) $x_1=6.74$ ) $P$ remains negative after the kink train enters the barrier, while for initial starting point ($x_1=29.05$ ) $\Delta P_{x_1}$ is enough great to change the $P$ into a positive value after arriving at the barrier. It is clear that for a location between these two values (where $V(\phi(x_1))=0$) there is no change in $P$ , as one can found in the Fig. \ref{AB}(a).

It should be noted that dislocations and impurities as spatial defects are not the only sources of perturbations. Dissipative effects, dispersion, different types of thermal noise and stochastic agents also strongly affect the properties and dynamical behaviour of periodic solutions. It is shown that thermally induced perturbations have definite impacts on the stability of some localized solutions of sG models \cite{1r}. In such studies the spatially-uniform Gaussian noisy force $\gamma(t)$ (with $<\gamma(t)>=0$) and delta-correlated Gaussian thermal noise $\gamma_{T}(x,t)$ (with    $<\gamma_{T}(x,t)>=0$ have been considered \cite{2r}. Generation of travelling waves in non-linear models in the presence of thermal noise is another important phenomenon related to the environmental effects \cite{3r, 4r}. The environmental perturbations as source of space-time dependent noise and random fluctuations are critical components of any non-linear and complex systems. Such defects play definitely important roles in classical and quantum description of physical systems, from microscopic to macroscopic scales \cite{5r,6r,7r,8r,9r,10r,11r}.  

\section{ soliton train-barrier interaction} \label{Field1}
The time evolution of soliton train solution during the interaction with medium disorders and impurities is an important issue due to its practical applications. The stability of the solution against medium defects can be studied by finding the emerged object after the interaction of the incident solution with a potential barrier.

\begin{figure}[ht]
\epsfxsize=9cm\centerline{\hspace{9cm}\epsfbox{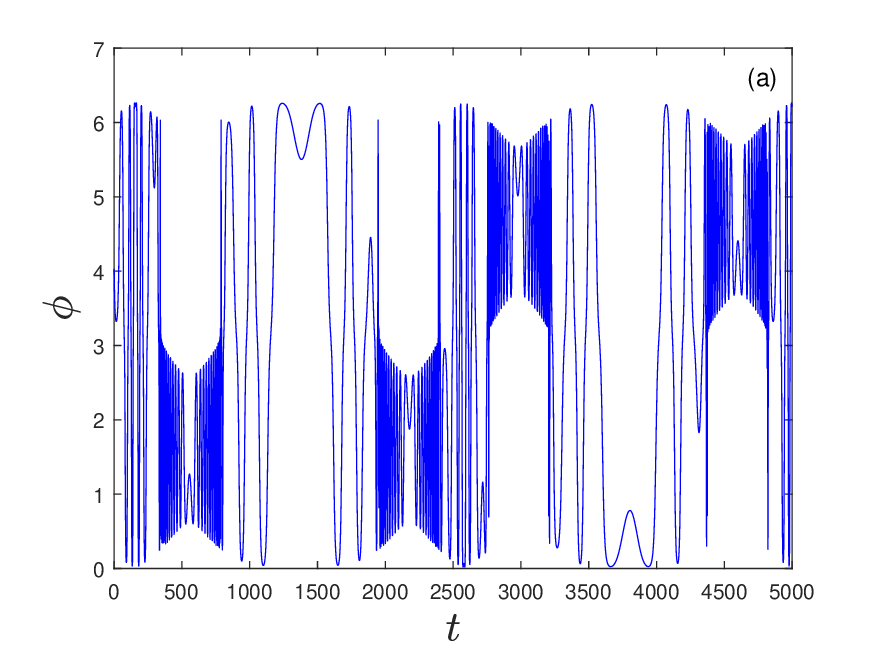}\epsfxsize=9cm\centerline{\hspace{-10cm}\epsfbox{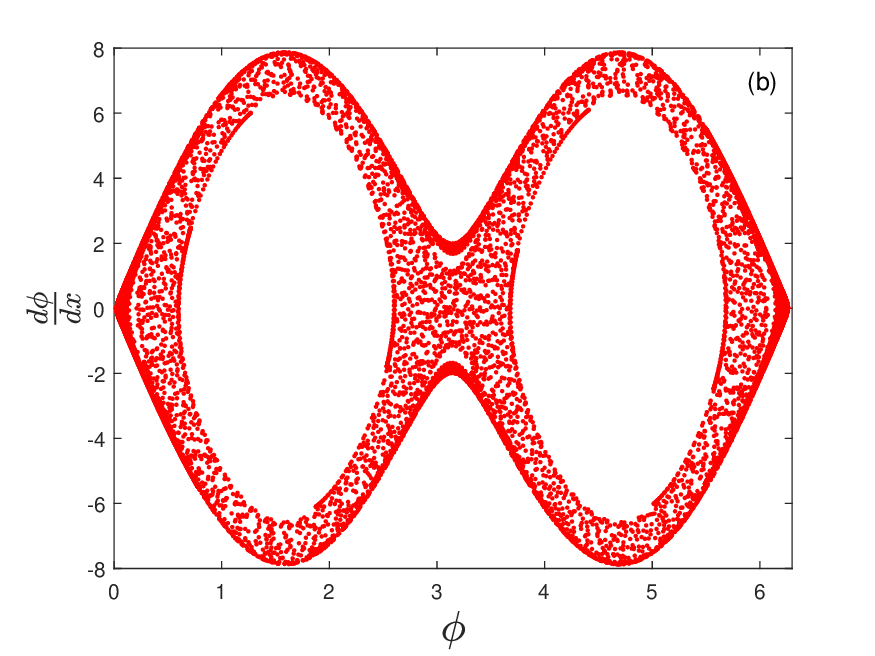}}}
\caption{(a) The $\phi$ diagram as a function of $t$ and (b) The phase diagram, with initial values of Fig. \ref{2}. \label{cl}}
\end{figure}

We have performed several  simulations to investigate the time evolution of a moving kink train after interaction with perturbation. Figure. \ref{cl}(a) demonstrates the time evolution of the field solution while interacting with a specified potential as a medium defect. The initial condition is the same as what has been chosen for Fig. \ref{2}(a). From this figure, we expect to observe moving $k\tilde{k}$ train interacting with the barrier. Therefore the initial conditions at the beginning point of perturbation (and thus the emerged solution from the barrier) are changing over time. As explained in section \ref{Field} the emerging solution is $kk$, $k\tilde{k}$, and $\tilde{k}\tilde{k}$ trains according to the initial condition. The positive slope (ascending part)  of the solution creates a $kk$ train of interaction,  while the negative slope (descending part) produces $\tilde{k}\tilde{k}$ configuration ( See Fig. \ref{AB}). It is clear that the flat parts of the emerging kink create $k\tilde{k}$ train. A complete set of initial conditions are created by positive, zero,  and negative slope parts. As one can find from the Fig. \ref{cl}, a sequence of $kk$, $k\tilde{k}$ and $\tilde{k}\tilde{k}$ trains are observed repeatedly, while their amplitudes are changing in time. 
  
The higher the kink slope as an initial condition, the faster the changes in the soliton amplitude after the interaction. Faster changes (compact parts) in Fig. \ref{cl}(a) relate to a higher (positive or negative) slope of the initial condition at the border of perturbation. The flat part of the initial condition provides a complete set of $k\tilde{k}$ solutions, as we can see within $1000<t<1500$, which depends on the velocity of the moving initial kink solution. 

Figure \ref{cl}(b) demonstrates the time evolution of the kink solution in phase space. As one can find, the phase space plot is formed by several times repeating the closed loop curve in the phase space (Fig. \ref{2}(b)), of course with some rapid changes  that are seen as jumps due to field-barrier interaction.  Figure \ref{cl}(a) shows that, the changes in the slope of $\phi$ are not uniform with respect to time. Sometimes the $\frac{d\phi}{dx}$ changes are sharp and sometimes slow. In figure \ref{cl}(b), the numerical data sets of figure \ref{cl}(a) have been ploted. The areas with high density of numerical points are related to the times when the slope of the field $\phi$ has more drastic changes. This figure contains important information about the stability of moving kink solutions while interacting with spatially  limited perturbations. 
We have examined several initial conditions. All our simulations have indicated that  limited amplitude barriers are not able to destroy the stability of soliton trains; which is an important outcome.

\begin{figure}[ht]
\epsfxsize=9cm\centerline{\hspace{9cm}\epsfbox{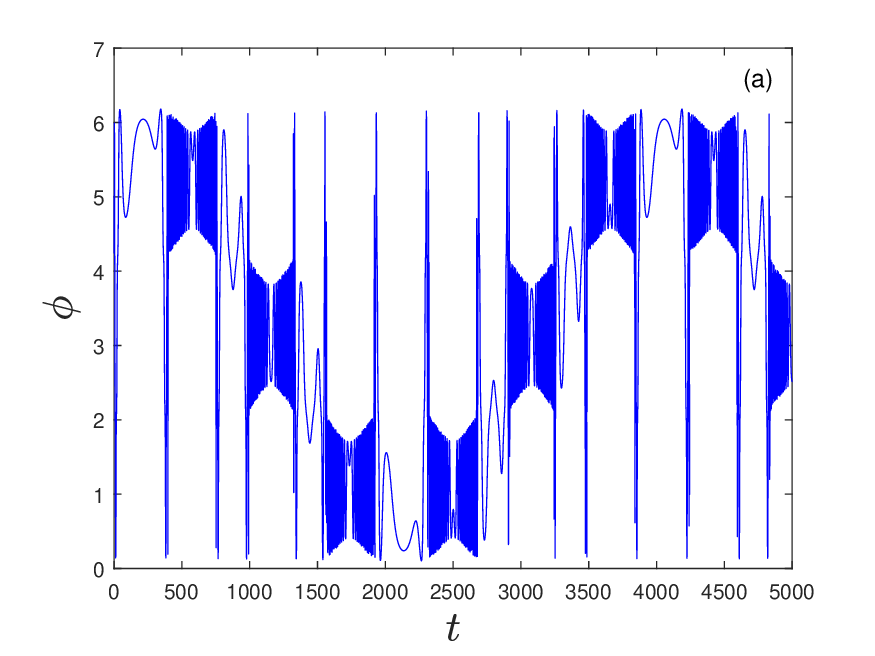}\epsfxsize=9cm\centerline{\hspace{-10cm}\epsfbox{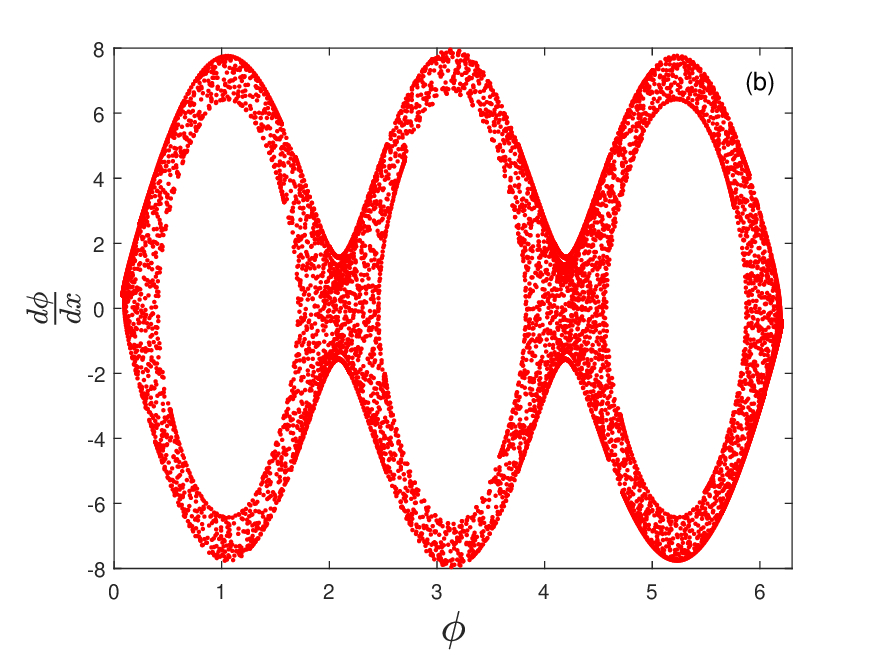}}}
\caption{(a) The $\phi$ diagram as a function of $t$ and (b) The phase
diagram, considering initial values of the solid curve in Fig. \ref{AB2}. \label{cl2}}
\end{figure}

Figure \ref{cl2} demonstrates field-potential interaction for $N=3$, the triple sG model. The general behaviour of interaction is the same as what we have found for the DsG model, except the generation of three sub-kinks in the profile of the field solution. We have not found different behaviour or instability condition for this model. Also, we have examined $N=4$ and $N=5$ solutions. No new phenomena have been observed. 

Time evolution of the field energy density $\rho(x,t)$ has been analysed during the interaction with medium perturbations too. It has been shown previously that, the total energy of single soliton changes during the interaction with potential \cite{k1, k2}. Recovery of Soliton energy into its initial quantity (before the interaction) depends on the nature of the soliton model and indeed the impact of the barrier. There is not much information about the time evolution of soliton train-barrier interactions. We have set up several simulations to examine the time evolution of the energy density for soliton chain solutions of $N=2$ and $N=3$ models while interacting with simple rectangular potentials of different amplitudes and widths. As expected, the total energy of the soliton train can be recovered with a good approximation if the barrier amplitude is sufficiently low. The final profile of the energy density after the interaction is a simultaneous function of the barrier amplitude and the state of the soliton at the initial and final points of the interaction. Thus, it is a complicated (and also challenging) study. In general, the energy of solitons, their sub-kink contents, amplitude, and widths are seriously affected by the interaction.  Figure \ref{2} presents the kink train-barrier interaction, where the barrier has greater density (higher $\epsilon$). The field stress density increases in the barrier, but it decreases again after finishing the field-potential interaction, when the kink train leaves the impurity area. The final value of the stress density after the interaction strongly depends on the field initial conditions and the characteristics of the barrier. The stability of the field energy and the shape of the kink train during and after the interaction with the impurity is understood by studying the time evolution of field potential energy and stress density. Figure \ref{cl} (\ref{cl2}) demonstrate interaction with the potential well (barrier) which determines by lower (higher) value for $\epsilon$. An important result that can be obtained from these figures is that the collision of the kink train with environmental perturbations may distort the shape of the periodic solution, but it does not have the ability to impose instability on the system.

\begin{figure}[ht]
\epsfxsize=20cm\centerline{\epsfbox{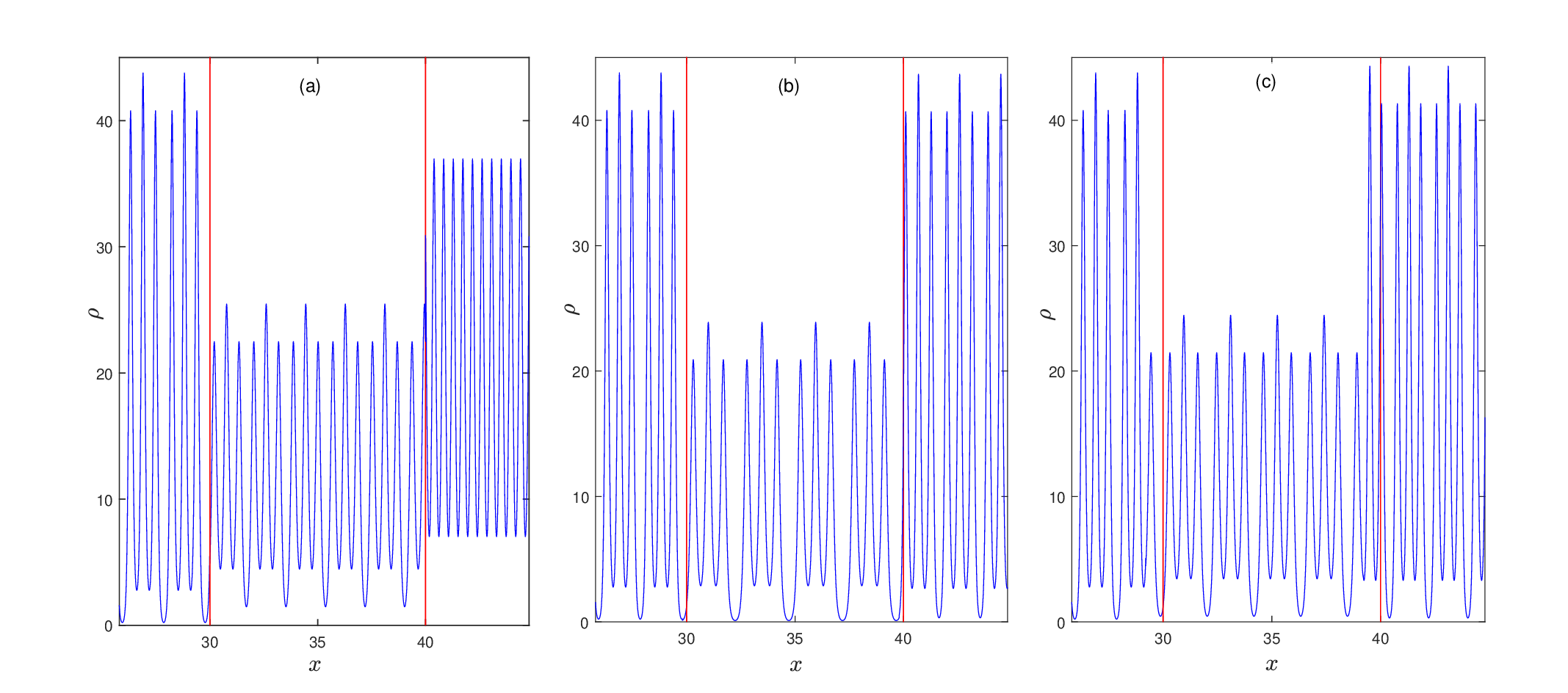}}
\caption{Profiles of the energy density as a function of scaled time ($t$) for $N=3$ and $P= -0.220000000000000$;   $\epsilon=10$ out of the interaction region and $\epsilon=5$ in the barrier; (a) $t=1$, (b) $t=1.2$ and (c) $t=1.8$ .\label{energy}}
\end{figure}

Figure \ref{energy} demonstrates three profiles of soliton-potential interaction at different times for a moving train with initial conditions of Fig. \ref{AB2}. The moving soliton collides with the potential, and as a result, the initial conditions for creating new kinks train inside the inhomogeneity region change with time. For this reason, the initial conditions also change at the endpoint of the potential region.

\section{concluding remarks}\label{Field3}

Impurities, dislocations, and other environmental inhomogeneities cause decisive effects on the behaviour of localized and quasi-periodic solutions of non-linear models. Such structural defects have been added to the model by different methods. We have investigated the effect of such thin barriers on the creation and evolution of multiple sine-Gordon kink trains by focusing on the dynamics of the double sine-Gordon and the triple sine-Gordon models. The analytical solution for soliton chain solutions (in series expansion) exists only for the double Sine-Gordon model. For this reason, we calculated the kink train solutions for multiple Sine-Gordon models using numerical calculations. We have checked the validity of numerical calculations by comparing our obtained results with analytical solutions of the double sine-Gordon model. We investigated the final configuration of kink train solutions after the initial kink chain-impurity interaction with details. The relationship between the final solution (also the kink train created in the impurity region) and the characteristics of the initial kink chain as well as the parameters of the potential barrier has been analysed. Our simulations show that regardless of the absolute location of the impurity and its width, the magnitude of the field and its slope at the start and end points of the interaction (the beginning and end of the perturbation region) are the most determining factors in shaping the final kink train solution after passing through the potential. The change in the value of model parameters ($\epsilon$ in our study) which is determined by the nature of the impurity, determines the amplitude and wavelength of final solutions after the interaction. The dynamics and general behaviour of interaction and the conditions for establishing the kink train solutions have been the same for all the investigated models ($N=2, ..., N=5$). The only difference is the appearance of sub-kinks, which originate from the natural properties of different models. 

The dynamics of soliton-potential interaction can not be fully understood without considering the internal mode effects. Energy transfer between the soliton (as a ground state) and  internal mode (as excited states) clearly changes the soliton's initial condition while interacting with medium impurities (as a potential barrier). Although the numerical solution automatically simulates the effects of internal modes, it does not show the effect of this phenomenon separately. For this reason, a suitable analytical model is necessary to understand the effect of internal modes on the dynamics of system evolution. Although we think that the effect is small, but this issue should be investigated in further works.

\acknowledgments{M. P. and K. J. acknowledge the support of
Ferdowsi University of Mashhad. N. R. acknowledges the support of Shahid Beheshti
University Research Council. }

\subparagraph*{Data Availability Statement}
This manuscript has no associated data or the data will not be deposited. [Authors' comment: This is a theoretical study and no experimental data has been listed.]


\end{document}